\documentclass[prl,twocolumn]{revtex4}%
\usepackage{amssymb}
\usepackage{graphicx}
\usepackage{amssymb}
\usepackage{graphicx}%

\begin{document}

\title{ A magnetic analog of the isotope\ effect in cuprates}
\author{ Rinat Ofer,$^{1}$ Galina Bazalitsky,$^{1}$ Amit Kanigel,$^{1}$ Amit
Keren$^{1}$, Assa Auerbach$^{1}$, James S. Lord$^{2}$, and Alex Amato$^{3}$}
\affiliation{{\normalsize {$^{1}$Physics Department, Technion-Israel Institute of
Technology, Haifa 32000, Israel}}}
\affiliation{{\normalsize {$^{2}$Rutherford Appleton Laboratory, Chilton Didcot,
Oxfordshire OX11 0QX, U.K.}}}
\affiliation{{\normalsize {$^{3}$Paul Scherrer Institute, CH 5232 Villigen PSI,
Switzerland}}}
\date{\today }

\begin{abstract}
We present extensive magnetic measurements of the (Ca$_{x}$La$_{1-x}$)(Ba$%
_{1.75-x}$La$_{0.25+x}$)Cu$_{3}$O$_{y}$ system with its four different
families ($x$) having a $T_{c}^{max}(x)$ variation of 28\% and minimal
structural changes. For each family, we measured the N\'{e}el temperature,
the anisotropies of the magnetic interactions, and the spin glass
temperature. Our results exhibit a universal relation $T_{c}=cJn_{s}$ for
all families, where $c\sim 1$, $J$ is the in plane Heisenberg exchange, and $%
n_{s}$ is the carrier density. This relates cuprate superconductivity to
magnetism in the same sense that phonon mediated superconductivity is
related to atomic mass.
\end{abstract}

\pacs{05.70.Ln, 74.40.+k, 74.25.Fy}
\maketitle

The critical temperature for superconductivity $T_{c}$ in metallic
superconductors varies with isotope substitution \cite{Maxwell}. This
observation, known as the isotope effect, played a key role in exposing
their mechanism for superconductivity. In contrast, the mechanism for
superconductivity in the cuprate is still elusive, but is believed to be of
magnetic origin \cite{Belivers}. Verifying this belief would require an
experiment similar to the isotope effect, namely, a measurement of $T_{c}$
versus the magnetic interaction strength $J$, with no other structural
changes in the compounds under investigation. Here we present such an
experiment using the (Ca$_{x}$La$_{1-x}$)(Ba$_{1.75-x}$La$_{0.25+x}$)Cu$_{3}$%
O$_{y}$ (CLBLCO) system with its four different superconducting families,
for which maximum $T_{c}$ ($T_{c}^{max}$) varies by 28\%. This is a large
change compared to Sn, which has the strongest isotope effect in nature
where $T_{c}$ varies only by 4\%. For each family, we measured the N\'{e}el
Temperature $T_{N}$ and the anisotropies of the magnetic interactions. This
allows us to obtain the Heisenberg coupling $J$. In addition, we determine
the spin glass temperature $T_{g}$ of underdoped samples. $J$, $T_{g}$ and $%
T_{c}$ allow us to generate a unified phase diagram for magnetism
and superconductivity from no doping to over doping. We combine
this result with a previous determination of \ the superconducting
carrier density $n_{s}$ \cite{KerenSSC03}, and demonstrate
experimentally a magnetic analog of the isotope effect.

CLBLCO is a high temperature superconductor (HTSC) system that belongs to
the YBa$_{2}$Cu$_{3}$O$_{7}$ (YBCO) group. Each value of $x$ in the range of
$0.1\leq x\leq 0.4$ \cite{preparation} is a family in the system. All
families are tetragonal. The difference in the unit cell parameters $a$ and $%
c/3$ between the two extreme families ($x=0.1$ and $0.4$) is 1\% \cite%
{preparation}. Therefore, variations in $T_{c}^{max}$ due to variation in
ionic radios are not relavent since they are expected to be on the order of $%
1~$K, and with opposite sign to observation \cite{ChenPRB05}. Moreover, the
level of disorder as detected by Ca NMR \cite{MarchandThesis} and Cu\ NQR
\cite{KerenToBe} is identical for the different families. The range of
accessible $y$'s in CLBLCO is so large that samples from the heavily
underdoped antiferromagnetic (AFM) parent compounds, to the
non-superconducting extreme overdoped, are obtainable. When varying $x$,
nearly concentric domes of $T_{c}$ versus oxygen doping $y$ are formed, with
the maximum value of $T_{c}$ varying from $58$~K at $x=0.1$ to $80$~K at $%
x=0.4$ \cite{Knizhnik}, as demonstrated by the open symbols in Fig.~\ref%
{phasediagram}. The magnetic properties of these compounds are
determined using the zero field muon spin relaxation ($\mu $SR)
technique.
\begin{figure}
[h]
\begin{center}
\includegraphics[
height=3.1in, width=3.7in
]%
{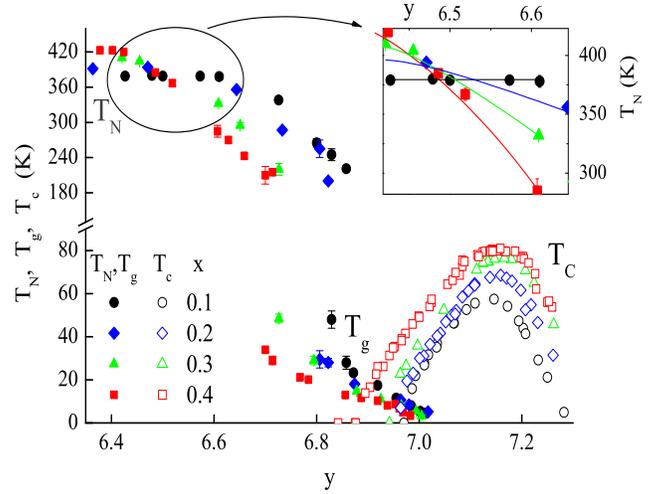}%
\caption{Phase diagram for the (CLBLCO)
(Ca$_{x}$La$_{1-x}$)(Ba$_{1.75-x}$La$_{0.25+x}$)Cu$_{3}$O$_{y}$
system.}%
\label{phasediagram}%
\end{center}
\end{figure}

\begin{figure}
[h]
\begin{center}
\includegraphics[
height=4.3in, width=3.5in
]%
{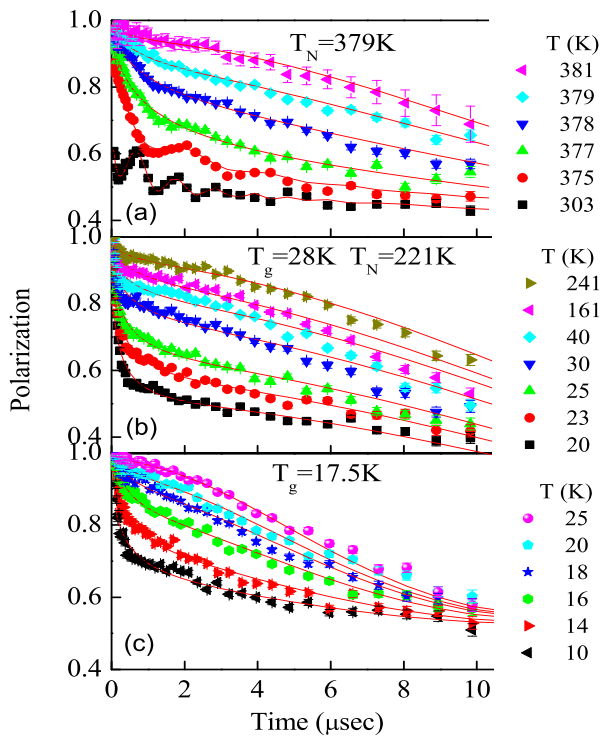}%
\caption{Time evolution of the muon polarization for (Ca$_{x}$La$_{1-x}$)(Ba$%
_{1.75-x} $La$_{0.25+x}$)Cu$_{3}$O$_{y}$ from the $x=0.1$ family
close to the magnetic critical temperature. (a) A sample with an
antiferromagnetic transition. (b) A sample with both an
antiferromagnetic and a spin-glass transition. (c) A sample with a
spin-glass transition. The solid lines are a fit to
Eq.(\protect\ref{fit}).}
\label{assgraph3}%
\end{center}
\end{figure}

Fig.~\ref{assgraph3} shows typical muon polarization $P(t)$
curves, at different temperatures, for three samples from the
$x=0.1$ family. At high temperatures the polarization curves from
all samples are typical of magnetic fields emanating from nuclear
magnetic moments. In this case the time dependence of the
polarization exhibits, as expected, a Gaussian decay. As the
temperature is lowered the sample enters a magnetic frozen phase
and the polarization relaxes much more rapidly. While the
transition from paramagnetic to the frozen state looks identical
for all samples, the behavior at very low $T$ is different and
indicates the nature of the ground state. Fig.~\ref{assgraph3}(a)
is an example of an antiferromagnetic ground state. When the
temperature decreases, long range magnetic order is established at
$\sim 377$~K reflected by spontaneous oscillations of the muon
polarization. Fig.~\ref{assgraph3}(c) is an example for a spin
glass (SG) transition at $\sim 17$~K. In this case the ground
state consists of magnetic islands with randomly frozen electronic
moments \cite{kanigel}, and consequently, the polarization shows
only rapid relaxation. When the transition is to a N\'{e}el or
spin glass state, the critical temperatures are named $T_{N}$ and
$T_{g}$, respectively. Fig~\ref{assgraph3}(b) presents an
intermediate case where the sample appears to have two
transitions. The first one starts below $240$~K, where the fast
decay in the polarization appears. Between $160$~K and $40$~K
there is hardly any change in the polarization decay, and at
$30$~K there is another transition manifested in a faster decaying
polarization. This behavior was observed in all the samples on the
border between antiferromagnet and spin-glass in the phase
diagram.

\begin{figure}
[h]
\begin{center}
\includegraphics[
height=3.1in, width=3.7in
]%
{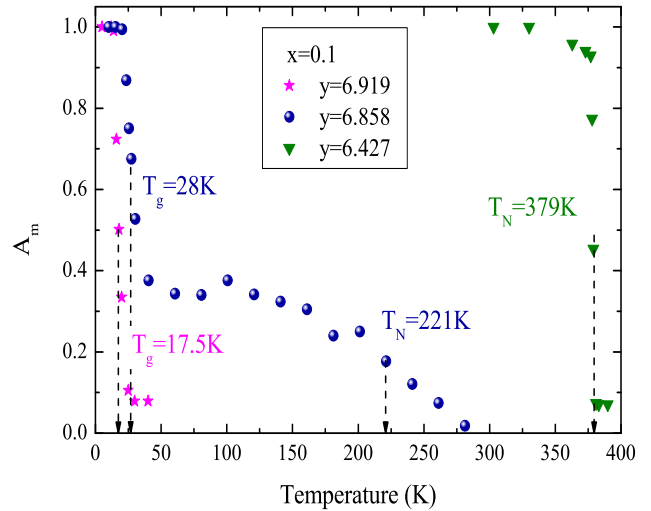}%
\caption{The magnetic volume fraction extracted from the muon
depolarization as a function of temperature for the 3 samples
shown in Fig.~\protect\ref{assgraph3}.}
\label{amT}%
\end{center}
\end{figure}

In order to determine the magnetic critical transition temperatures, the
data were fitted to a sum of two functions: a Gaussian, and a rapidly
relaxing function that describes the magnetic field due to frozen electronic
moments, namely,
\begin{equation}
P(t)=A_{n}\exp (\frac{-\Delta ^{2}t^{2}}{2})+ \label{fit}
\end{equation}%
\begin{eqnarray}
+A_{m}\left\{ a\exp (-\sqrt{ \lambda _{1}t})+(1-a)\exp
(-\sqrt{\lambda _{2}t})\cos (\omega t)\right\}.\nonumber
\end{eqnarray}%

In this function $A_{n}$ and $A_{m}$ represent the amplitudes (i.e. the
volume fraction) of the nuclear and magnetic parts, respectively, and $%
\lambda _{1},\lambda _{2}$ are the relaxation rates of the magnetic part. In
the fit $\Delta $ is determined from high temperatures, and the sum $%
A_{m}+A_{n}=1$ is constant at all temperatures. The solid lines in Fig.~\ref%
{assgraph3} are the fits of Eq.(\ref{fit}) to the data. Fig. \ref{amT} shows
$A_{m}$ as a function of temperature, for the three samples in Fig.~\ref%
{assgraph3}. Above the transition, where only nuclear moments contribute, $%
A_{m}$ is close to zero. As the temperature decreases, the frozen magnetic
part increases and so does $A_{m}$, at the expense of $A_{n}$. For the pure
AFM and SG phases, the transition temperature was determined as the
temperature at which $A_{m}$ is half of the saturation value. For the
samples with two transitions, two temperatures were determined using the
same principle. The coexistence of N\'{e}el and spin-glass states indicates
that the transition from one kind of ground state to the other is a first
order quantum phase transition.

The full phase diagram of the CLBLCO compound, obtained from the
measurements described above, including the $T_{g}$ measurements in the
superconducting state from Ref.~\cite{kanigel}, is presented in Fig.~\ref%
{phasediagram}. From this diagram it is clear that, at high doping levels,
CLBLCO is a pure superconductor. As the oxygen doping decreases considerably
from optimum, these compounds can be in a mixed state of magnetic islands,
which freeze at $T_{g}$, and superconductivity in between. Upon further
decrease in doping, the system has a pure spin glass ground state. Even
further underdoping towards the parent compounds leads to abrupt replacement
of the glassy state with a long range antiferromagnetic order with $%
T_{N}\sim 400~$K, which seems to saturate. The inset of Fig.~\ref%
{phasediagram} \ shows that $y\sim 6.5$ is a crossing point, and
at higher doping, families with high $x$ values have lower $T_{N}$
and $T_{g}$. \ In contrast, at doping levels below $y=6.5$,
families with higher $x$ values (and higher $T_{c}^{max}$) have
the higher $T_{N}$. This is the first indication of correlation
between $T_{c}^{max}$ and magnetic interactions.
\begin{figure}
[h]
\begin{center}
\includegraphics[
height=3.1in, width=3.6in
]%
{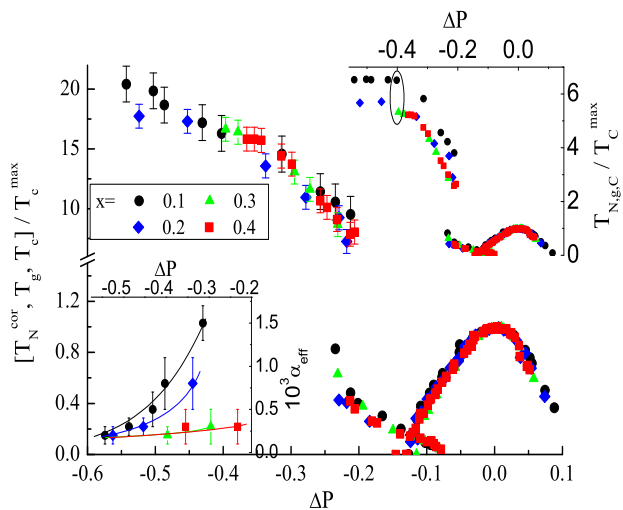}%
\caption{Upper inset: the phase diagram after the scaling
described in the text. Lower inset: The effective anisotropies as
a function of $\Delta P$ for the different families, see text for
details. The solid lines are guides to the eye. Main figure: phase
diagram after both scaling and extraction of the contributions
from anisotropies as described in the text. $T_{N}^{cor}=J$ for
the parent compounds.}
\label{scaling2}%
\end{center}
\end{figure}

In order to untangle the phase diagram, we use the scaling relation
introduced in Ref.~\cite{kanigel} that caused both $T_{c}$ and $T_{g}$ data
of many cuprate families to collapse into a single curve \cite{KerenPRB03}.
This means plotting, for each family, $T_{c}$ divided by $T_{c}^{max}$ of
the family, as a function of $\Delta P$ where $\Delta P=K(x)\Delta y$, $%
\Delta y=y-y_{0}$ is the chemical doping measured from optimum $y_{0}$, and $%
K(x)$ is the scaling parameter determined for each family so that
all the superconducting domes collapse onto a single curve
\cite{kanigel}. We use $K=0.77, 0.67, 0.54, 0.47$, and $y_0=7.135,
7.15, 7.155, 7.15$ for the x=0.1 to 0.4 respectivly. The results
are presented in the inset of Fig.~\ref{scaling2}. We find that,
with the exception of the $x=0.1$ family in the AFM region, the
scaling relation holds perfectly well for the entire phase
diagram.

\begin{figure}
[h]
\begin{center}
\includegraphics[
height=3in, width=3.6in
]%
{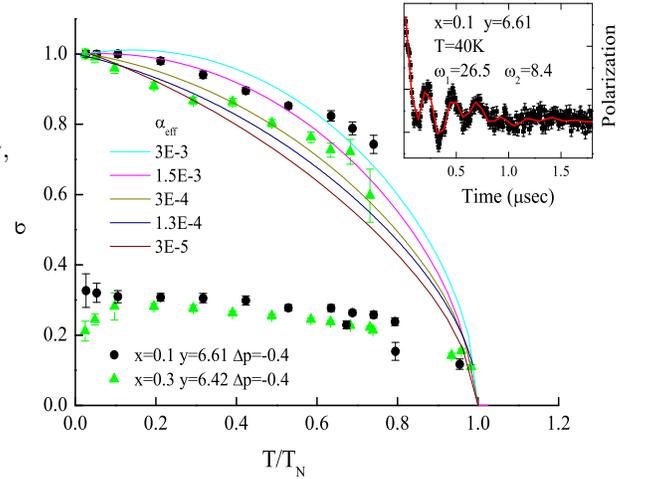}
 \caption{The normalized staggered magnetization as a
function of the normalized temperature. The symbols are the
experimental results, taken by measuring the oscillation frequency
of the polarization curves. The solid lines are the theoretical
curves plotted according to Eqs. \protect\ref{SBMF1} to
\protect\ref{SBMF4}. Inset: an example of muon polarization for a
sample in the N\'{e}el state at $T\rightarrow 0$, the solid line
is a fit to Eq.(\protect\ref{oscillation fit})}
\label{alpha2}%
\end{center}
\end{figure}

Nevertheless, we would like to check whether the scaling relation can be
extended to the $x=0.1$ family. We suspected that in $x=0.1$ anisotropies
might be different from the other families. It is well established that a
pure 2D AFM orders magnetically only at $T=0$, and that $T_{N}$ for 3D AFM
is finite. Intermediate cases are described by anisotropic interactions
where $J$ \ and $\alpha J$ are the in and out of plane coupling
respectively. For small $\alpha $ the reduction of the magnetic order
parameter $M$ with increasing $T$ is fast so that at $\alpha =0$ the 2D
limit is recovered. A plot the normalized order parameter $\sigma =M/M_{0}$,
where $M_{0}$ is the order parameter at $T\rightarrow 0$, versus $T/T_{N}$
should connect (1,0) to (0,1) (see Fig.~\ref{alpha2}) in a way that is
determined only by $\alpha $. At the same time $J\propto T_{N}\ln (\alpha )$
(up to $\ln $ of $\ln $ of $\alpha $ corrections) \cite{KeimerPRB92} so that
knowledge of $\alpha $ would lead to $J$.

To test our hypothesis we measure $P(t)$ with high timing resolution at $%
T\rightarrow 0$, and extract $\sigma $. A typical time dependent
polarization is depicted in the inset of Fig.~\ref{alpha2}. The best fit of
the polarization is achieved with the function
\begin{equation}
P(t)=\sum_{i=1}^{3}A_{i}\exp (-\lambda _{i}t)\cos (\omega _{i}t)
\label{oscillation fit}
\end{equation}%
with $\omega _{3}=0$, which is shown in the inset~by the solid line. The
reason for multiple frequencies is that the muons stop at different sites in
the unit cell. Since the muon oscillation frequency is proportional to the
local magnetic field it experiences, $\omega (T)/\omega _{0}$ is equivalent
to $\sigma $. The order parameter extracted from the high frequency, around
few tens of MHz ($\omega _{0}\sim 27$~MHz in our case) is known to agree
with neutron scattering determination of $\sigma $ \cite{KerenPRB93}. The
lower frequency is believed to emerge from metastable muon sites and is not
used for further analysis.

In Fig.~\ref{alpha2} we present $\sigma $ for the two different CLBLCO
samples with $x=0.1$ and $0.3$ having the same $\Delta P$ that is marked in
the inset of Fig.~\ref{scaling2}. The data sets starting at $\sigma =1$ is
from the high frequency and will be used for comparison with theory. Clearly
the reduction of the magnetization with increasing temperatures is not the
same for these two samples, and therefore their anisotropies are different.
Since $\sigma $ is less sensitive to increasing $T$ in the $x=0.1$ family
than in the $x=0.3$ family the $\alpha $ of $x=0.1$ must be larger.
Consequently $T_{N}$ of the $x=0.1$ \textquotedblleft turns
out\textquotedblright\ to be too high due to $\alpha $ and not $J$. This
could explain the deviation from scaling of the $x=0.1$ family.

To account for the anisotropies quantitatively we assume that CLBLCO could
be considered as a 2D magnet with weak anisotropies since the chain layers
are partially full with oxygen even for the parent compounds. Therefore we
use the Hamiltonian
\begin{equation}
H=J(\sum_{i,\delta _{\Vert }}\mathbf{S}{{_{i}}\cdot \mathbf{S}{_{i+\delta
_{\Vert }}}}+\alpha _{xy}\sum_{i,\delta _{\Vert }}{S_{i}^{z}S_{i+\delta
_{\Vert }}^{z}}+\alpha _{\perp }\sum_{i,\delta _{\perp }}\mathbf{S}{%
_{i}\cdot \mathbf{S}_{i+\delta _{\perp }}})  \label{Ham}
\end{equation}%
where $\delta _{\Vert }$ and $\delta _{\perp }$ are the in and out of plane
neighbor spacings, respectively. We apply the self-consistent
Schwinger-boson mean-field (SBMF) theory \cite{ArovasPRB98} to calculate $%
\sigma (\alpha _{eff},t)$ where $t=T/J$, $\alpha _{eff}=z_{xy}\alpha
_{xy}+z_{\perp }\alpha _{\perp }$, and the $z$'s are the number of
neighbors. The calculation is done by solving simultaneously for every $%
\alpha _{eff}$ and $t$ two equations: a self consistency equation%
\begin{equation}
h=2\alpha _{eff}\left[ 1-2K(\Delta ,h,t)\right]   \label{SBMF1}
\end{equation}%
and a constraint equation ensuring one Schwinger-boson per site
\begin{equation}
K(\Delta ,h,t)+K(\Delta ,0,t)=1.  \label{SBMF2}
\end{equation}%
In these equations
\begin{equation}
K(\Delta ,h,t)=2.32\int_{0}^{1}\frac{(1+\Delta +h)}{\omega (\Delta ,h,\gamma
)}\left[ n(\omega (\Delta ,h,\gamma ),t)+1/2\right] \rho (\gamma )d\gamma ,
\label{SBMF3}
\end{equation}%
the density of states is given by
\begin{equation}
\rho (\gamma )=\frac{2}{\pi ^{2}} \int\limits_{0}^{1}\left[
(1-t^{2})(1-t^{2}+\gamma^{2}t^{2})\right] ^{-1/2}dt, \label{SBMF4}
\end{equation}
or its approximation \cite{MathHandBook}, $\omega (\Delta
,h,\gamma )=2.32((1+\Delta +h)^{2}-\gamma ^{2})^{1/2}$, and
$n(\omega ,t)=\left[ \exp (\omega /t)-1\right] ^{-1}$. Finally,
$\sigma (\alpha _{eff},t)=h(\alpha _{eff},t)/h(\alpha _{eff},0)$.
$t_{N}$ and $T_{N}$ are defined by $\sigma (\alpha
_{eff},t_{N})=0$ and $T_{N}=Jt_{N}$. The theoretical order
parameter for several $\alpha _{eff}$ as a function of $T/T_{N}$
is also presented in Fig.~\ref{alpha2} by the solid lines.

When comparing theory and experiment we focus on the low
temperature data, up to $200$~K, where the theory is most
accurate. We determine the best $\alpha _{eff}$ and present them
in the lower inset of Fig.~\ref{scaling2}. The solid lines are
guides to the eye. Using the corresponding $t_{N}(\alpha_{eff})$
we obtain the corrected $T_{N}$ defined as
\begin{equation}
T_{N}^{cor}\equiv T_{N}/t_{N}(\alpha _{eff})  \label{keimerTn}
\end{equation}%
where $T_{N}^{cor}=J$ for the parent compounds. At $\Delta P=-0.4$ where $%
T_{N}$ of all samples is saturated and they can be compared we find $%
J=950(60)$, $1130(70)$, $1260(80)$, and $1330(80)$~K for $x=0.1$, $0.2$, $0.3
$, and $0.4$ respectivly.

This variation of $J$ is probably due to an increasing buckling angle and
decreasing lattice constants with increasing $x$. Neutron diffraction show
that the buckling angle of the optimally doped samples changes by $1.3$
degrees from $x=0.1$ to $x=0.4$. Similarly the lattice constant $a$ changes
by $0.03$ \AA\ \cite{ChmaissemNature99}. According to Ref.~\cite%
{ChndramouliAJC03}, where $J$ is plotted as a function of the Cu-O-Cu
buckling angle and bond length for a variety of materials, this king of
changes can produce a 30\% variation in $J$.

Determining $t_{N}\ $experimentally for the AFM compounds allows us to
present a modified phase diagram using $T_{N}^{cor}$. The modified phase
diagram showing $T_{N}^{cor},T_{g}$ and $T_{c}$ normalized by $T_{c}^{\max }$
versus $\Delta P$ is presented in Fig.~\ref{scaling2}. Note, $T_{g}$ is not
corrected since it is determined by interactions between islands, and
anisotropies are not expected to play an important role in this case. The
large error bars for $T_{N}^{cor}$ is due to the pure determination of $%
t_{N}(\alpha _{eff})$. Nevertheless, the scaling now works for all four
families, with $T_{c}^{max}$ variation of 28\%, in most of the AFM region as
well. This demonstrates that an energy scale $J_{f}$ , unique for each
family but varying between families, controls both $T_{N}$, $T_{g}$ and $%
T_{c}$.

In a previous work it was found that CLBLCO obeys the Uemura relation $%
T_{c}\propto n_{s}$ \cite{Uemura}, where $n_{s}$ is the superconducting
carrier density, in both under- and overdoped regions \cite{KerenSSC03}.
Combining this fact with our present finding that a magnetic energy scale
controls $T_{c}^{\max }$, suggests the formula
\begin{equation}
T_{c}=cJ_{f}n_{s}(\Delta P)  \label{magnetop}
\end{equation}%
where the constant $c=0.7(1)$ (using the definition $n_{s}(0)=0.08$ carrier
per Cu), $\Delta P=0$ at optimal doping for all families, and $n_{s}$ is
family independent.

Equation (\ref{magnetop}) is a magnetic equivalent of the isotope effect $%
T_{c}=cM^{-1/2}$. We demonstrate this magnetic effect for 15\% variation of $%
T_{c}^{\max }$ without any theory. This is far greater than the $T_{c}$
variation due to the strongest isotope effect in metallic superconductors.
Using the SBMF theory we extend the effect further to 28\% of $T_{c}^{\max }$%
. Therefore, our data is a strong support to the belief that magnons are
responsible for producing pairing in the cuprates.

We would like to thank the ISIS pulsed muon facility at Rutherford Appleton
Laboratory, UK and the S$\mu $S facility at Paul Scherrer Institute,
Switzerland. This work was funded by the Israeli Science Foundation and the
Posnansky Research Fund in High Temperature superconductivity.

\end{document}